\documentclass[twocolumn]{article}

\usepackage{multicol,booktabs,float}
\usepackage{graphicx,multirow}
\usepackage{amsmath,amssymb,amsfonts}
\usepackage{xcolor,colortbl}
\usepackage{url,siunitx,comment}
\usepackage[utf8]{inputenc}
\usepackage[normalem]{ulem}
\usepackage[compact]{titlesec}
\usepackage[margin=1.085in]{geometry}

\title{Corona Games: Masks, Social Distancing and Mechanism Design}

\author{Balázs Pejó, Gergely Biczok\\
	Laboratory of Cryptography and System Security \\
	Deptartment of Networked Systems and Services \\
	Budapest University of Technology and Economics\\
	email: \{familyname\}@crysys.hu}

\date{}

\begin{document}
	
\maketitle
	
	\begin{abstract}
		Pandemic response is a complex affair. Most governments employ a set of quasi-standard measures to fight COVID-19 including wearing masks, social distancing, virus testing and contact tracing. We argue that some non-trivial factors behind the varying effectiveness of these measures are selfish decision-making and the differing national implementations of the response mechanism.\\
		In this paper, through simple games, we show the effect of individual incentives on the decisions made with respect to wearing masks and social distancing, and how these may result in a sub-optimal outcome. We also demonstrate the responsibility of national authorities in designing these games properly regarding the chosen policies and their influence on the preferred outcome. We promote a mechanism design approach: it is in the best interest of every government to carefully balance social good and response costs when implementing their respective pandemic response mechanism.
	\end{abstract}
		
	\section{Introduction}
	
	The current coronavirus pandemic is pushing individuals, businesses and governments to the limit. People suffer owing to restricted mobility, social life and income, complete business sectors face an almost 100\% drop in revenue, and governments are scrambling to find out when and how to impose and remove restrictions. In fact, COVID-19 has turned the whole planet into a ``living lab'' for human and social behavior where feedback on response measures employed is only delayed by around two weeks (the incubation period). From the 24/7 media coverage, all of us have been introduced to a set of quasi-standard measures introduced by national and local authorities, including wearing masks, social distancing, virus testing, contact tracing and so on. It is also clear that different countries have had different levels of success employing these measures as evidenced by the varying normalized death tolls and confirmed cases\footnote{Johns Hopkins Coronavirus Resource Center. \url{https://coronavirus.jhu.edu/map.html}}. 
	
	We believe that apart from the intuitive (e.g., genetic differences, medical infrastructure availability, hesitancy, etc.), there are two significant factors that have not received enough attention. First, the \emph{individual incentives} of citizens, e.g., ``is it worth more for me to stay home than to meet my friend?'', have a significant say in every decision situation. While some of those incentives can be inherent to personality type, clearly, there is a non-negligible rational aspect to it, where individuals are looking to maximize their own utility. Second, countries have differed in their specific \emph{implementation} of response measures, e.g., whether they have been distributing free masks (affecting the efficacy of mask wearing in case of equipment shortage) or providing extra unemployment benefits (affecting the likelihood of proper self-imposed social distancing). Framing pandemic response as a mechanism design problem, i.e., architecting a complex response mechanism with a preferred outcome in mind, can shed light on these factors; what's more, it has the potential to help authorities (mechanism designers) fight the pandemic efficiently. The objective of this paper is to show that both individual incentives and the actual design and implementation of the holistic pandemic response mechanism can have a major effect on how this pandemic plays out.
	
	\subsubsection*{Contribution. }
	
	In this paper we model decision situations during a pandemic with game theory where participants are rational, and the proper design of the games could be the difference between life and death. Our main contribution is two-fold. First, regarding decisions on wearing a mask, we show that i) the equilibrium outcome is not socially optimal under full information, ii) when the status of the players are unknown the equilibrium is not to wear a mask for a wide range of parameters, and iii) when facing an infectious player it is almost always optimal to wear a mask even with low protection efficiency. Furthermore, for social distancing, using current COVID-19 statistics we showed that i) going out is only rational when it corresponds to either a huge benefit or staying home results in a significant loss, and ii) we determined the optimal duration and meeting size of such an out-of-home activity. 
	Second, we take a look at pandemic response from a mechanism design perspective, and demonstrate that i) different government policies influence the outcome of these games profoundly, and ii) individual response measures (sub-mechanisms) are interdependent. Specifically, we discuss how contact tracing enables targeted testing which in turn reduces the uncertainty in individual decision making regarding both social distancing and wearing masks. We recommend governments treat pandemic response as a mechanism design problem when weighing response costs vs. the social good.
	
	\subsubsection*{Organization. }
	
	The remaining of the paper is structured as follows. Section~\ref{sec:related} briefly describes related work while Section~\ref{sec:pre} recaps some basics of game theory. Section~\ref{sec:mask} develops and analyzes the Mask Game adding uncertainty, mask efficiency and multiple players to the basic model. Section~\ref{sec:sd} develops and analyzes the Distancing Game including the effects of meeting duration and size. Section~\ref{sec:mech} frames pandemic response as a mechanism design problem using the design of the two games previously introduced as examples. Finally, Section~\ref{sec:con} outlines future work and concludes the paper.
	
	\section{Related Work}
	\label{sec:related}
	
	In this section we review some well-known epidemic spreading models and game-theoretic works in relation to pandemics.
	
	COVID-19 have been modelled using different models: for instance using SIR \cite{carletti2020covid}, SEIQR \cite{zhang2020impact}, and SIDARTHE \cite{giordano2020modelling}. Which model suits the ongoing epidemic best is still undetermined. Besides the model, the input data instantiating the model may be imperfect as well, thus some efforts are also made to account for potential inaccuracies in the reported data \cite{hong2020estimation}. An orthogonal extension of these models is proposed in \cite{santosh2020covid}, which discusses how factors such as hospital capacity, test capacity, demographics, population density, vulnerable people and income could be integrated into these models. In contrast with the previous models, the one in \cite{lagos2020games} takes into consideration the networked structure of human interconnections and the locality of interactions, without attempting a mean-field approach. In the following we briefly review some related research efforts in the intersection of epidemics and game theory. For a comprehensive survey we refer the reader to \cite{chang2019game}.
	
	Some researchers modeled the behavioral changes of people to a pandemic: for instance in \cite{poletti2012risk} authors used evolutionary game theory, and showed that slightly reducing the number of people an individual was in contact with could make a difference regarding the spread of disease. Another group showed that there was a critical level of concern, i.e., empathy, by the infected individuals above which the disease is eradicated rapidly \cite{eksin2017disease}. Others focused on the mobility habits of people traveling between areas affected unevenly by the disease, and found conflict between the Nash Equilibrium (individually optimal strategy) and the Social Optimum (optimal group strategy) only under specific changes in economic and epidemiological conditions \cite{zhao2018strategic}. In \cite{bairagi2020controlling} an optimization problem was formalized by accommodating both isolation (modeled by how far individuals are from home) and social distancing (how far individuals are from each other). Authors also provided incentives for maintaining social distancing to prevent the spread of COVID-19 (i.e., making ``staying home'' the Nash Equilibrium). Moreover, social distancing was also shown to be able to delay the epidemic until a vaccine becomes widely available \cite{reluga2010game}. 
	
	Several studies focused on how the availability of vaccines affects human behaviour. A model was introduced in \cite{bhattacharyya2011wait} where vaccine delayers relied on herd immunity and vaccine safety information generated by early vaccinators. Consequently, the Nash Equilibrium was ``wait and see''. Another study concerning this vaccination dilemma proposed a model with incentives for individuals to choose the prevention strategy according to risks and expenses in the epidemic campaign \cite{bauch2004vaccination}. Similarly, researchers in \cite{van2008self} showed the optimal use of anti-viral treatment by individuals when they took into account the direct and indirect costs of treatment. The game-theoretic model in \cite{sun2009selfish} focused on the various level of drug stockpiles in different countries, and found controversial results: sometimes there was an optimal solution with a central planner (such as the WHO), which improved on the decentralized equilibrium, but other times the central planner's solution (minimizing the number of infected persons globally) required some countries to sacrifice part of their population. 
	
	The exact dynamics of demand and supply for medical resources at different phases of a pandemic was also studied \cite{chen2020pandemic}. Predicting such dynamics would provide a quantitative basis for mechanism designers (e.g., decision makers of healthcare systems) to understand the potential imbalance of supply and demand. The authors extended the concepts of reserving and capital management in the classical insurance literature and aimed to provide a quantitative framework for quantifying and assessing pandemic risk, and developed optimal strategies for stockpiling spatio-temporal resources.
	
	The Centers for Disease Control and Prevention created a policy review of social distancing measures for pandemic influenza in non-healthcare settings \cite{fong2020nonpharmaceutical}. They identified measures to reduce community influenza transmission such as isolating the sick, tracing contacts, quarantining exposed people, closing down school, changing workplace habits, avoiding crowds, and restricting movement. The impact of several of these (and wearing masks) was studied in \cite{silva2020covid} in which the authors model the pandemic by emulating people, business and government. Other researchers demonstrated that early school and workplace closure, and restriction of international travel are independently associated with reduced national COVID-19 mortality \cite{papadopoulos2020impact}. On the other hand, lock-down procedures could have devastating impact on the economy. This was studied in \cite{chao2020simplified} with a modified SIR model and time-dependent infection rate. The authors found that, surprisingly, in spite of the economic cost of the loss of workforce and incurred medical expenses, the optimum point for the entire course of the pandemic is to keep the strict lock-down as long as possible.
	
	As detailed above, related work has mostly studied narrowly focused specifics of epidemic modelling such as the intricate behaviour of individuals in relation with vaccines or the preferred actions of mechanism designers such as healthcare system operators. In contrast, our work takes a step back, and focuses on the big picture: we model decision situations during a pandemic as games with rational participants, and promote the proper design of these games. We highlight the responsibility of mechanism designers such as national authorities in constructing these games properly with adequately chosen policies, taking into account their interdependent nature. 
	
	\section{Preliminaries}
	\label{sec:pre}
	
	In this section we shortly elaborate on the main game theoretical notions used in this paper, to enable the conceptual understanding of the implications of our results. 
	
	Game theory \cite{harsanyi1988general} is ``the study of mathematical models of conflict between intelligent, rational decision-makers''. Almost every multi-party interaction can be modeled as a game. In relation to COVID-19, decision makers could be individuals (e.g., whether to wear a mask), cities (e.g., whether to enforce wide-range testing within the city), governments (e.g., whether to apply contact tracking within the country), or companies (e.g., whether to apply social distancing within the workplace). Potential decisions are referred to as strategies; decision makers (players) choose their strategies rationally so as to maximize their own utility.
	
	The Nash Equilibrium (NE) --- arguably the most famous solution concept --- is a set of strategies where each player's strategy is a best response strategy. This means every player makes the best/optimal decision for itself as long as the others' choices remain unchanged. NE provides a way of predicting what will happen if several entities are making decisions at the same time where the outcome also depends on the decisions of the others. The existence of a NE means that no player will gain more by unilaterally changing its strategy at this unique state. Another game-theoretic concept is the Social Optimum, which is a set of strategies that maximizes social welfare. Note, that despite the fact that no one can do better by changing strategy, NEs are not necessarily Social Optima (we refer the reader to the famous example of the Prisoner's Dilemma \cite{harsanyi1988general}). In fact, it is a central problem in game theory how much a distributed outcome (NE) is worse than a centrally planed social optimum.
	
	If one knows the NE they prefer as the outcome of a game, e.g., everybody wearing a mask, and they have the power to instantiate the game accordingly, i.e., fixing the structure, game flow and any free parameters, then we talk about mechanism design~\cite{mas1995microeconomic}. In a way, mechanism design is the inverse of game theory; although a significant share of efforts within this field deals with auctions, mechanism design is a much broader term widely applicable to any mechanism, e.g., optimal organ matching for transplantation or school-student allocation, aimed at achieving a given steady state result. 
	
	\section{The Mask Game}
	\label{sec:mask}
	
	Probably the most visible consequence of COVID-19 are masks: before their usage was mostly limited to some Asian countries, hospitals, constructions and banks (in case of a robbery). Due to the coronavirus pandemic, an unprecedented spreading of mask-wearing can be seen around the globe. Policies have been implemented to enforce their usage in some places, but in general, it has been up to the individuals to decide whether to wear a mask or not based on their own risk assessment. In this section, we model this decision situation via game theory. We assume that there are several types of masks, providing different level of protection. 
	
	\begin{itemize}
		\item \textbf{No} Mask corresponds to the behavior of using no masks during the COVID-19 (or any) pandemic. Its cost is consequently 0; however, it does not offer any protection against the virus. 
		\item \textbf{Out} Mask is the most widely used mask (e.g., cloth mask or surgical mask). They are meant to protect the environment of the individual using it. They work by filtering out droplets when coughing, sneezing or simply talking, therefore they limit the spreading of the virus. They do not protect the wearer itself against an airborne virus. The cost of deciding for this protection type is noted as $C_{out}>0$.
		\item \textbf{In} Mask is the most protective prevention gear designed for medical professionals (e.g., FFP2 or FFP3 mask with valves). Valves make it easier to wear the mask for a sustained period of time, and prevent condensation inside the mask. They filter out airborne viruses while breathing in, however the valved design means they do not filter the while air breathing out. Note that CDC guidelines\footnote{Centers for Disease Control and Prevention. \url{https://www.cdc.gov/coronavirus/2019-ncov/prevent-getting-sick/prevention.html}} recommend using a cloth/surgical mask for the general public, while valved masks are only recommended for medical personnel in direct contact with infected individuals. The cost of this protection type is $C_{in}>>C_{out}$.
	\end{itemize}
	
	Besides which mask they use (i.e., the available strategies), the players are either susceptible or infected\footnote{We simplify the well-known SIR model \cite{diekmann2000mathematical} since in case of COVID-19 it is currently unclear if and for how long an individual is resistant after recovery.}. The latter has some undesired consequence; hence, we model it by adding a cost $C_i$ to these players' utility (which is magnitudes higher than even $C_{in}$, i.e., $C_i>>C_{in}>>C_{out}$). We summarize all the parameters and variables used for the Mask game in Table \ref{tab:parMASK}. 
	
	\begin{table}[h]
		\centering
		\begin{tabular}{c|l}
			Variable & Meaning \\
			\hline
			$C_{out}$ & Cost of playing \textbf{out} \\
			$C_{in}$ & Cost of playing \textbf{in} \\
			$C_i$ & Cost of being infected \\
			\hline
			$C_{use}$ & Cost of playing \textbf{use} \\
			$\rho$ & Prob. of being infected \\
			$p$ & Prob. of using a mask \\
			\hline
			$a$ & Protection Efficiency \\
			$b$ & Spreading Efficiency \\
		\end{tabular}
		\vspace{0.1cm}
		\caption{Parameters of the Mask Games}
		\label{tab:parMASK}
	\end{table}
	
	Using these states and masks, we can present the basic game's payoffs where two players with known health status meet and decide which mask to use. The payoff matrix in Table \ref{tab:mask_game_payoffBOTH} corresponds to the case when both players are susceptible. Note, that in case both players are infected, the payoff matrix would be the same with an additional constant $C_i$. Table \ref{tab:mask_game_payoffONE} corresponds to the case when one player is infected while the other is susceptible. 
	
	\begin{table}[ht]
		\centering
		\begin{tabular}{c|ccc}
			& no & out & in\\
			\hline
			no & $(0,0)$ & $(0,C_{out})$ & $(0,C_{in})$ \\
			out & $(C_{out},0)$ & $(C_{out},C_{out})$ & $(C_{out},C_{in})$ \\
			in & $(C_{in},0)$ & $(C_{in},C_{out})$ & $(C_{in},C_{in})$ \\
		\end{tabular}
		\vspace{0.1cm}
		\caption{Payoff matrices when both players are susceptible}
		\label{tab:mask_game_payoffBOTH}
	\end{table}
	
	In Table \ref{tab:mask_game_payoffBOTH} it is visible that both players' cost is minimal when they do not use any masks, i.e., the Nash Equilibrium of the game when both players are susceptible is (\textbf{no}, \textbf{no}). This is also the social optimum, meaning that the players' aggregated cost is minimal. The same holds in case both players are infected, as this only adds a constant $C_i$ to the payoff matrix. 
	
	\begin{table*}[ht]
		\centering
		\begin{tabular}{c|ccc}
			& no & out & in\\
			\hline
			no & $(C_i,C_i)$ & $(0,C_{out}+C_i)$ & $(C_i,C_{in}+C_i)$ \\
			out & $(C_{out}+C_i,C_i)$ & $(C_{out},C_{out}+C_i)$ & $(C_{out}+C_i,C_{in}+C_i)$ \\
			in & $(C_{in},C_i)$ & $(C_{in},C_{out}+C_i)$ & $(C_{in},C_{in}+C_i)$ \\
		\end{tabular}
		\vspace{0.1cm}
		\caption{Payoff matrices for the cases when only one player is susceptible.}
		\label{tab:mask_game_payoffONE}
	\end{table*}
	
	When only one of the players is susceptible as represented in Table \ref{tab:mask_game_payoffONE}, using no mask is a dominant strategy for the infected player\footnote{Note that the payoffs does not take into account the legal consequences of a deliberate infection such as in \url{https://www.theverge.com/2020/4/7/21211992/coughing-coronavirus-arrest-hiv-public-health-safety-crime-spread}. }, since it is a best response, independently of the susceptible player's action. Consequently, the best option for the susceptible player is \textbf{in}, i.e., the NE is (\textbf{in}, \textbf{no}). On the other hand, the social optimum is different: (\textbf{no},\textbf{out}) would incur the least burden on the society since $C_{out}<<C_{in}$.
	
	In social optimum, susceptible players would benefit, through a positive externality, from an action that would impose a cost on infected players; therefore it is not a likely outcome. In fact, such a setting is common in man-made distributed systems, especially in the context of cybersecurity. A well-fitting parallel is defense against Distributed Denial of Service Attacks (DDoS) attacks~\cite{khouzani2013}: although it would be much more efficient to filter malicious traffic at the source (i.e., \textbf{out}), Internet Service Providers rather filter at the target (i.e., \textbf{in}) owing to a rational fear of free-riding by others. 
	
	\subsection{Bayesian Game}
	
	Since in the basic game no player plays \textbf{out}, we simplify the choice of the players to either \textbf{use} a mask or \textbf{no} (hence, we note the cost of a mask with $C_{use}$). To represent the situation more realistically, we introduce ambiguity about the status of the players: we denote the probability of being infected as $\rho$. We know from the basic game that if both players are infected (with probability $\rho^2$) or susceptible (with probability $(1-\rho)^2$) they play (\textbf{no,no}), while if only one of them is infected (with probability $2\cdot\rho\cdot(1-\rho)$) the infected player plays \textbf{no}, while the susceptible plays \textbf{use}. Hence, the players play \textbf{no} in most of the cases (e.g., with probability $1-(\rho\cdot(1-\rho))$).
	
	On the other hand, this is not the case if we do not assume that the players know their statuses. Consequently, with uncertainty we must minimize the costs of the players: if both players are infected with equal probability, the payoff for Player 2 is Equation \ref{eq:util_bayes} where $p_n$ is the probability that Player n plays \textbf{use} (otherwise she plays \textbf{no}). The payoff for the other player is similar since the game is symmetric. In more detail, the first two lines correspond to the case when Player 2 is not infected (hence the multiplication with $1-\rho$ at the beginning), while the last line captures when she is infected. Either way, she plays \textbf{use} with probability $p_2$, which incurs a cost of $C_{use}$. Otherwise she plays \textbf{no}, which has no cost except when Player 1 is infected and she plays \textbf{no} as well. 
	
	\begin{equation}
		\begin{split}
			\label{eq:util_bayes}
			U_2=(1-\rho)\cdot[(1-\rho)\cdot[p_2\cdot C_{use}+(1-p_2)\cdot0]+\\
			\rho\cdot[p_2\cdot C_{use}+(1-p_2)\cdot[(1-p_1)\cdot C_i+p_1\cdot0]]]+\\
			\rho\cdot[p_2\cdot(C_{i}+C_{use})+(1-p_2)\cdot C_i]
		\end{split}
	\end{equation}
	
	Since this formula is linear in $p_2$, its extreme point within [0,1] is situated exactly at the boundary. We take its derivative to uncover the function steepness: the condition for the function to be decreasing (i.e., higher probability for using a mask corresponds to lower cost) is seen below. Consequently, the only scenario which might admit wearing a mask with non-zero probability corresponds to the availability of sufficiently cheap masks.
	
	\begin{equation}
		\label{eq:partial}
		\frac{\partial U_2}{\partial p_2}<0\Leftrightarrow \frac{C_{use}}{C_i}<\rho\cdot(1-\rho)\cdot(1-p_1)\le1
	\end{equation}
	
	\subsubsection*{Example. }
	
	Lets assume Alice is going to meet Bob after a long time without any correspondence. Consequently, she does not know whether Bob has been exposed to SARS-CoV-2 recently. Actually, Alice herself could have been exposed as well without her knowledge, as up to $80\%$ of the infectious cases could be asymptomatic.\footnote{Centre for Evidence-Based Medicine. \url{https://www.cebm.net/covid-19/covid-19-what-proportion-are-asymptomatic/}} For this reason, without taking into account any available spatial data, she estimates that they could be infectious with $\rho=50\%$: either yes or no. She also does not have any information about Bob's mask wearing habits, so she guesses $p_1=0.5$ as well. 
	
	Alice is tested at her workplace every day, and she is sent to a 1-week quarantine without payment if tested positive. If we represent Alice as an average American, she earns approximately $1000$ USD per week\footnote{Bureau of Labour Statistics. \url{https://www.bls.gov/news.release/pdf/wkyeng.pdf}}, hence, we set $C_i=1000$. Substituting these into the right side of Equation (\ref{eq:partial}), she decides to wear a mask only if it costs less than $125$ USD, which does hold as of September 2020. 
	
	\subsection{Efficiency Game}
	
	In the basic game we assumed \textbf{in} provides perfect protection from infected players, while \textbf{out} protects the other player fully. However, in real life, these strategies only mitigate the infection by decreasing its probability  (i.e., $\rho$) to some extent. For this reason, we define $a, b\in[0,1]$ in a way that the smaller value of the parameter corresponds to better protection; $a$ measures the protection efficiency of the protection strategy, while $b$ captures the efficiency of eliminating the further spread of the disease. Consequently, $a$ and $b$ was set in the previous cases to $a_\textbf{out}=0,a_\textbf{in}=1$ (\textbf{in} prevents further spreading, while \textbf{out} does not), $b_\textbf{out}=1$ (\textbf{out} has no effect on protecting the player) $b_\textbf{in}=0$ (\textbf{in} perfectly protects the player). 
	
	We simplify the action space of the players as we did in the Bayesian game: \textbf{in} and \textbf{out} is merged into \textbf{use} Obviously, \textbf{no} corresponds to $a_{no}=b_{no}=0$. We abuse the notion $a$ and $b$ to represent $a_{use}$ and $b_{use}$ respectively. We set $b=\frac23$, as surgical masks on the infectious person reduce cold \& flu viruses in aerosols by 70\% according to \cite{milton2013influenza}. Parameter $a$ is much harder to measure. It should be $a\le b$ since any mask keeps the virus inside the players more efficiently than stopping the wearer from getting infected. For the sake of the analysis we set $a=\frac{b}{2}=\frac13$, but any other choice would be possible.
	
	We are interested in the mask-wearing probability of a susceptible player when the other player is infected.\footnote{The Bayesian game combined with efficiency is left for future work due to the lack of space.} The utility in such a situation is shown in Equation (\ref{eq:util_eff}), where for simplification we defined $p=p_1=p_2$, i.e., both players play a specific strategy with the same probability. With such a constraint, we restrict ourselves from finding all the solutions; however, since the game is symmetric, an equilibrium of this reduced game is also an equilibrium when the players could use different strategy distributions.
	
	\begin{equation}
		\begin{split}
			\label{eq:util_eff}
			U=&p^2\cdot(C_{use}+C_i\cdot a\cdot b)+\\
			&p\cdot(1-p)\cdot(C_{use}+C_i\cdot a)+\\
			&(1-p)\cdot p\cdot(C_i\cdot b)+\\
			&(1-p)^2\cdot C_i\\
			\Rightarrow U=&p^2\cdot(C_{use}+C_i\cdot 0.\dot{2})+\\
			&p\cdot(1-p)\cdot(C_{use}+C_i)+\\
			&(1-p)^2\cdot C_i
		\end{split}
	\end{equation}
	
	From this we easily deduce that \textbf{use} corresponds to a smaller cost that \textbf{no} if $\frac{C_{use}}{C_i}<\frac79$, which holds by default as $C_{use}\ll C_i$ (even for less efficient masks). Moreover, \textbf{use} (i.e., $p=1$) is the best response most of the time because of the following.
	
	\begin{enumerate}
		\item The utility is a second order polynomial, hence it has one extreme point.
		\item This extreme point is a minimum due to $U^{\prime\prime}=\frac49\cdot C_i>0$.
		\item The utility (i.e., cost) is decreasing on the left and increasing on the right of this minimum point.
		\item The utility's minimum point is at $p=\frac94\cdot\frac{C_i-C_{use}}{C_i}$ due to $U^\prime=C_{use}-C_i+\frac49\cdot C_i\cdot p$.
		\item The minimum point is expected to be above 1 due to $C_{use}\ll C_i$.
		\item $p\in[0,1]$ is on the left of the minimum point, hence, a higher $p$ corresponds to a smaller cost. 
	\end{enumerate}

	\subsection{Multi-Player Game}
	
	This game can be further extended by allowing more players to participate. In this extension --- if we assume all players meet with probability 1 --- with any number of infected players (who play \textbf{no} as we showed already) all the susceptible players should play \textbf{in}. This NE is the SO as well if the ratio of the infected (which is identical to the probability $\rho$ of being infected) is sufficiently high: the accumulated cost when the susceptible players play \textbf{in} (and the infected play \textbf{no}) is less than the accumulated cost when the infected players play \textbf{out} (and the susceptible play \textbf{no}) if $\frac{C_{in}}{C_{out}}<\frac{\rho}{1-\rho}$. Although it is mathematically possible that the infected plays \textbf{no} in the SO, but it is doubtful: both the cost of \textbf{in} is significantly higher than \textbf{out}, and the infection ratio $\rho$ is low (at least at the beginning of the pandemic).
	
	\section{The Distancing Game}
	\label{sec:sd}
	
	Another phenomenon most people has experienced during the current COVID-19 pandemic is social distancing. Here we introduce a simple Distancing Game; it is to be played in sequence with the previously introduced Mask Game: once a player decided to meet up with friends via the Distancing Game, she can decide whether to wear a mask for the meeting playing the Mask Game. To improve readability, we summarize all the corresponding parameters and variables in Table \ref{tab:parDIST}. 
	
	\begin{table}[ht]
		\centering
		\begin{tabular}{c|l}
			Variable & Meaning \\
			\hline
			$C$ & Cost of staying home\\
			$B$ & Benefit of going out \\
			\hline
			$m$ & Mortality rate \\
			$L$ & Value of Life\\
			\hline
			$\rho$ & Probability of infection \\
			$p$ & Probability of meeting \\
			\hline
			$t$ & Time duration of meeting \\
			$g$ & Group size of meeting \\
		\end{tabular}
		\vspace{0.1cm}
		\caption{Parameters of the Distancing Games}
		\label{tab:parDIST}
	\end{table}
	
	We represent the cost of getting infected with $m\cdot L$, i.e., the mortality rate of the disease multiplied with the player's evaluation about her own life.\footnote{This is an optimistic approximation, as besides dying the infection could bear other tolls on a player.} Besides the risk of getting infected, going out or attending a meeting could benefit the player, denoted as $B$. On the other hand, staying home or missing a meeting could have some additional costs, denoted as $C$. The probability of getting infected is denoted as $\rho$. With these notations, the utility of the Distancing Game is captured on the left of Equation (\ref{eq:dist}), where $p$ is the probability of going out. Since this is linear in $p$, its maximum is either at $p=0$ (stay home) or $p=1$ (go out). Precisely, the player prefers to stay home if the right side of Equation (\ref{eq:dist}) holds. 
	
	\begin{equation}
		\label{eq:dist}
		U=p\cdot(B-\rho\cdot m\cdot L)-(1-p)\cdot C \hspace{1cm} \frac{B+C}{\rho\cdot m}<L
	\end{equation}
	
	\subsubsection*{Example. }
	
	For instance, should a rational American citizen (e.g., Alice) go out based on how much she values her life? We estimate\footnote{Data from \url{https://www.worldometers.info/coronavirus/} (accessed September 2020)} $m=0.034$ and $\rho=0.0077$ as $0.028\approx\frac{\#\{\text{deceased}\}}{\#\{\text{all cases}\}}<m<\frac{\#\{\text{deceased}\}}{\#\{\text{closed cases}\}}\approx0.04$ while $\frac{\#\{\text{active cases}\}}{\#\{\text{population}\}}\approx0.0077$. 
	
	Using these values, Alice should go out only if she values her life less than $3820(=\frac1{0.034\cdot0.0077})$ times the benefit (of going out) and the loss (of staying home) together. According to \cite{trottenberg2013guidance}, the value of a statistical life in the US was 9.2 million USD in 2013, which is equivalent to 11.3 million USD in 2020 (with $0.3\%$ interest rate). This means, Alice should only meet someone if the benefit of the meeting (and thus the cost of missing out) would amount to more than USD $2,958$ ($=\frac{11.3M}{3820}$).
	
	\begin{figure*}[bt]
		\centering
		\includegraphics[width=0.9\textwidth]{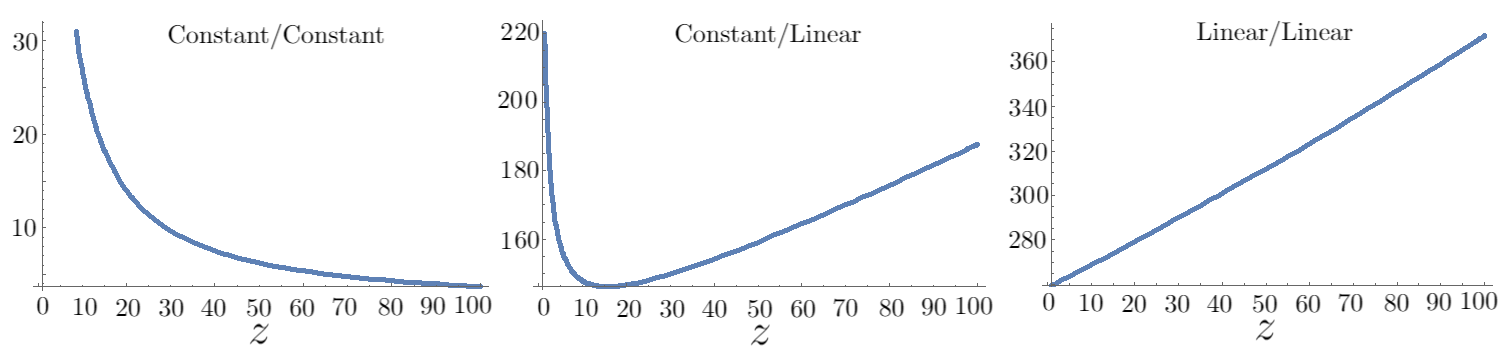}
		\caption{A few examples for various benefit and cost functions of the lower limit on the life value which would ensure that Alice (i.e., a rational American) would stay home (i.e.,left side of Equation (\ref{eq:dist_ext})) with $m=0.034$ and $\rho=0.0077$. }
		\label{fig:dist_ext}
	\end{figure*}

	\subsection{Number of Participants and Duration}
	
	One way to improve the above model is by introducing meeting duration and size. Leaving our disinfected home during a pandemic is risky, and this risk grows with the time. Similarly, a meeting is riskier when there are multiple participants involved. In the original model, we captured the infection probability with $\rho=1-(1-\rho)$. This ratio increases to $1-(1-\rho)^{g\cdot t}$ when there are $g$ possible infectious sources for $t$ time. Since $g$ and $t$ are interchangeable, we merge this two together under a common notation: $z=g\cdot t$. 
	
	This extended model can be used to determine the optimal duration and size of a meeting, once the player decided to go out according to the basic Distancing Game. We define $0<z<100$, as no player has infinite time or meeting partners. Moreover, the benefit and the loss of attending and missing a meeting should depend on this new parameter. For instance, staying home in isolation for a longer period might cause anxiety, which could get worse over time (i.e., increasing the cost); on the other hand, attending a meeting with many friends at the same time could significantly boost the experience (i.e., increase the benefit). Consequently, a rational person should leave her home only if Equation (\ref{eq:dist_ext}) holds which is the extension of the right side of Equation (\ref{eq:dist}). 
	
	\begin{equation}
		\label{eq:dist_ext}
		\max_{0<z<100}\left(\frac{B(z)+C(z)}{(1-(1-\rho)^z)\cdot m}\right)<L
	\end{equation}
	
	In Figure \ref{fig:dist_ext}, we present three use-cases of the formula inside the maximization above: the left one represents the case when both the benefit and the cost are constant, the right one corresponds to the case when both of them are linear. In the middle, there is a mixture of these two. Note that we needed to restrict $z$ to be under a certain amount as it represents both the time and the size of a meeting. 
	
	\section{Pandemic Mechanism Design}
	\label{sec:mech}
	
	\begin{figure*}[tb]
		\centering
		\includegraphics[width=0.9\textwidth]{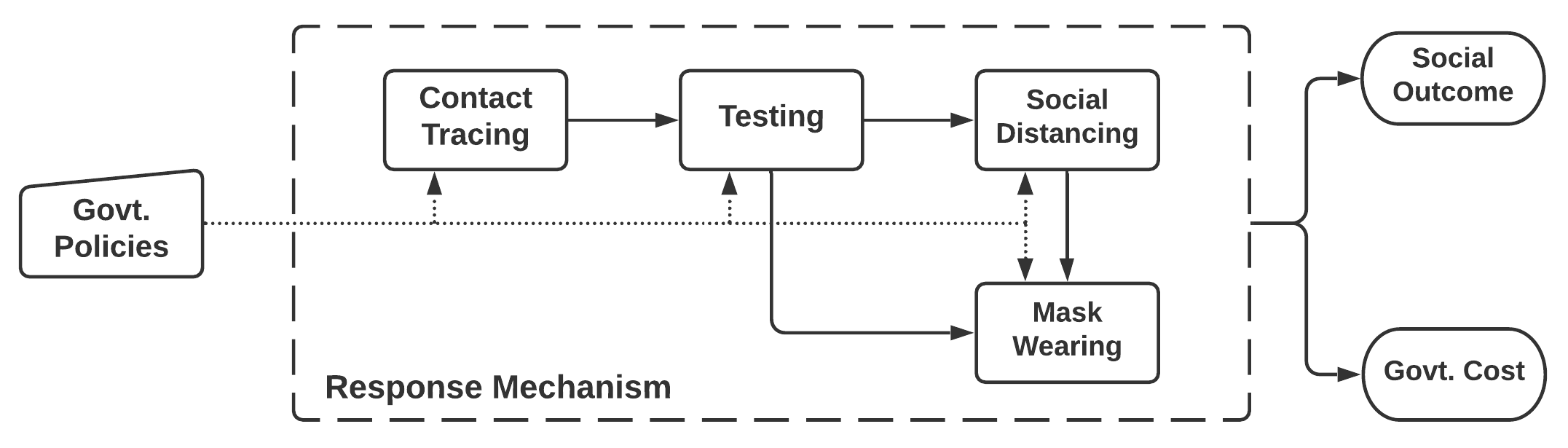}
		\caption{Pandemic response mechanism as influenced by government policy (dotted lines) and the interplay of sub-mechanisms (solid lines)}
		\label{fig:bigpicture}
	\end{figure*}
	
	Pandemic response is a complex affair. The two games described above model only parts of the bigger picture.
	
	\subsection{The government as mechanism designer}
	
	We refer to the collection (and interplay) of measures implemented by a specific government fighting the epidemic in their respective country as \emph{mechanism}. Consequently, decisions made with regard to this mechanism constitutes \emph{mechanism design}~\cite{mas1995microeconomic}. In its broader interpretation, mechanism design theory seeks to study mechanisms achieving a particular preferred outcome. Desirable outcomes are usually optimal either from a social aspect or maximizing a different objective function of the designer.
	
	In the context of the corona pandemic, the immediate response mechanism is composed of e.g., wearing a mask, social distancing, testing and contact tracing, among others. Note that this is not an exhaustive list: financial aid, creating extra jobs to accommodate people who have just lost their jobs, declaring a national emergency and many other conceptual vessels can be utilized as sub-mechanisms by the mechanism designer, i.e., usually, the government; we do not discuss all of these in detail due to the lack of space. Instead, we shed light on how government policy can affect the sub-mechanisms, how sub-mechanisms can affect each other and, finally, the outcome of the  mechanism itself. We illustrate the importance of mechanism design applying different policies to our two games, and adding testing and contact tracing to the mix.
	
	\subsection{Policy impact on sub-mechanisms and the final mechanism}
	
	Here we analyze the impact of commonly seen policies: compulsory mask wearing, distributing free masks, limiting the amount of people gathering and total lock-down.
	
	\subsubsection*{Compulsory mask wearing and free masks. }
	
	If the government declares that wearing a simple mask is mandatory in public spaces (such as shops, mass transit, etc.), it can enforce an outcome (\textbf{out},\textbf{out}) that is indeed socially better than the NE. The resulting strategy profile is still not SO, but it i) allocates costs equally among citizens; ii) works well under the uncertainty of one's health status; and iii) may decrease the first-order need for large-scale testing, which in turn reduces the response cost of the government. By distributing free masks, the government can reduce the effect of selfishness and, potentially, help citizens who cannot buy or afford masks owing to supply shortage or unemployment.
	
	\subsubsection*{Limiting the amount of people gathering and total lock-down. }
	
	If the government imposes an upper limit $l$ for the size of congregations, this will put a strict upper bound on the ``optimal meeting size'' $g^*$, and the resulting group size will be $\min(l,g^*)$. Note that if $l < g^*$ then it creates an ``opportunity'' for longer meetings (larger $t$), as Equation (\ref{eq:dist_ext}) maximizes for $z = g t$. On the other hand, if the chosen restrictive measure is a total lock-down, both the Distancing Game and the Mask Game are rendered moot, as people are not allowed to leave their households.
	
	\subsubsection*{Testing and contact tracing. }
	
	It is clear that the Distancing and the Mask Games are not played in isolation: people deciding to meet up invoke the decision situation on mask wearing. On the other hand, so far we have largely ignored two other widespread pandemic response measures: testing and contact tracing.
	
	With appropriately designed and administered coronavirus tests, medical personnel can determine two distinct features of the tested individual: i) whether she is actively infected spreading the virus and ii) whether she has already had the virus, even if there were no or weak symptoms. (Note that detecting these two features require different types of tests, able to show the presence of either the virus RNA or specific antibodies, respectively.) In general, testing enables both the tested person and the authorities to make more informed decisions. Putting this into the context of our games, testing i) reduces the uncertainty in Bayesian decision making, 
	and ii) enables the government to impose mandatory quarantine thereby removing infected players. Even more impactful, mandatory testing (as in Wuhan\footnote{New York Times. \url{https://www.nytimes.com/2020/05/26/world/asia/coronavirus-wuhan-tests.html}}) completely eliminates the Bayesian aspect, essentially rendering the situation to a full information game: it serves as an exogenous ``health oracle'' imposing no monetary cost on the players. To sum it up, the testing sub-mechanism outputs results that serve as inputs to both the Distancing and the Mask Game.
	
	Naturally, a ``health oracle'' does not exist: someone has to bear the costs of testing. From the government's perspective, mandatory mass testing is extremely expensive\footnote{But not without precedence, e.g., in Slovakia (\url{https://edition.cnn.com/world/live-news/coronavirus-pandemic-10-18-20-intl/h_beb93495fe9b83701023eafd5f28e39d})}. (Similarly, from the concerned individual's perspective, a single test could be unaffordable.) Contact tracing, whether traditional or mobile app-based, serves as an important input sub-mechanism to testing~\cite{ferretti2020quantifying}. It identifies the individuals who are \emph{likely} affected based on spatial proximity, and inform both them and the authorities about this fact. In game-theoretic terms, for such players, the benefit of testing outweigh the cost (per capita) with high probability. From the mechanism designer's point of view, contact tracing reduces the overall testing cost by enabling \emph{targeted testing}, potentially by orders of magnitude, without sacrificing proper control of the pandemic. Another potential cost of contact tracing for individuals could be the loss of privacy. Note that mobile OS manufacturers are working on integrating privacy-preserving contact tracing into their platform to eliminate adoption costs for installing an app\footnote{Apple. \url{https://covid19.apple.com/contacttracing}}.
	
	\subsubsection*{The big picture. }
	
	As far as pandemic response goes, the mechanism designer has the power to design and parametrize the games that citizens are playing, taking into account that sub-mechanisms affect each other. After games have been played and outcomes have been determined, the cost for the mechanism designer itself are realized (see Figure~\ref{fig:bigpicture}). This cost function is very complex incorporating factors from ICU beds through civil unrest to a drop in GDP over multiple time scales~\cite{mcdonald2008macroeconomic}. Therefore, governments have to carefully balance the -- very directly interpreted -- social optimum and their own costs; this indeed requires a mechanism design mindset.
	
	\section{Conclusion}
	\label{sec:con}
	
	In this paper we have made a case for treating pandemic response as a mechanism design problem. Through simple games modeling interacting selfish individuals we have shown that it is necessary to take individual incentives into account during a pandemic. We have also demonstrated that specific government policies significantly influence the outcome of these games, and how different response measures (sub-mechanisms) are interdependent. As an example we have discussed how contact tracing enables targeted testing which in turn reduces the uncertainty from individual decision making regarding social distancing and wearing masks. Governments have significantly more power than traditional mechanism designers in distributed systems; therefore it is even more crucial for them to carefully study the tradeoff between social good and the cost of the designer when implementing their pandemic response mechanism.
	
	\subsubsection*{Limitations and future work. }
	
	Clearly, we have just scratched the surface of pandemic mechanism design. The models presented are simple and mostly used for demonstrative purposes. Also, the mechanism design considerations are only quasi-quantitative without proper formal mathematical treatment. In turn, this gives us plenty of opportunity for future work. A potential avenue is extending our models to capture the temporal aspect, combine them with epidemic models as games played by many agents on social graphs, and parametrize them with real data from the ongoing pandemic (policy changes, mobility data, price fluctuations, etc.). Relaxing the rational decision-making aspect is another prominent direction: behavioral modeling with respect to obedience, other-regarding preferences and risk-taking could be incorporated into the games. Finally, a formal treatment of the mechanism design problem constitutes important future work, incorporating hierarchical designers (WHO, EU, nations, municipality, household), an elaborate cost model, and analyzing optimal policies for different time horizons. If done with care, these steps would help create an extensible mechanism design framework that can aid decision makers in pandemic response.
	
	\bibliographystyle{apalike}
	\bibliography{ref}
\end{document}